\documentclass[12pt,a4paper]{article}

\usepackage{amsmath,amssymb,amsthm}
\usepackage[matrix,arrow]{xy}

\newcommand{\A}{\mathcal{A}}        
\renewcommand{\AA}{\mathbb{A}}      
\renewcommand{\a}{\alpha}           
\newcommand{\B}{\mathcal{B}}        
\newcommand{\ket}[1]{|#1\rangle}    
\newcommand{\braCket}[3]{\langle#1\mathbin|#2\mathbin|#3\rangle}
\newcommand{\C}{\mathbb{C}}         
\newcommand{\D}{\mathcal{D}}        
\newcommand{\del}{\partial}         
\newcommand{\dl}{\delta}            
\DeclareMathOperator{\dom}{dom}     
\newcommand{\eps}{\epsilon}         
\newcommand{\F}{\mathcal{F}}        
\newcommand{\FF}{\mathbb{F}}        
\newcommand{\Ga}{\Gamma}            
\newcommand{\g}{\mathfrak{g}}       
\newcommand{\ga}{\gamma}            
\newcommand{\gl}{\mathfrak{gl}}     
\renewcommand{\H}{\mathcal{H}}      
\newcommand{\id}{{\rm id}}          
\renewcommand{\L}{\mathcal{L}}      
\newcommand{\La}{\Lambda}           
\newcommand{\la}{\lambda}           
\newcommand{\N}{\mathbb{N}}         
\newcommand{\Om}{\Omega}            
\newcommand{\om}{\omega}            
\newcommand{\omf}{\tilde\omega}     
\newcommand{\op}{\oplus}            
\newcommand{\owl}{\overline}        
\newcommand{\ox}{\otimes}           
\DeclareMathOperator{\ran}{ran}     
\newcommand{\Sf}{\mathbb{S}}        
\newcommand{\sepword}[1]{\quad\hbox{#1}\quad} 
\newcommand{\set}[1]{\{\,#1\,\}}    
\DeclareMathOperator{\spec}{spec}   
\DeclareMathOperator{\tr}{tr}       
\renewcommand{\th}{\theta}          
\newcommand{\thalf}{\tfrac{1}{2}}   
\newcommand{\tihalf}{\tfrac{i}{2}}  
\newcommand{\tquarter}{\tfrac{1}{4}} 
\newcommand{\tiquarter}{\tfrac{i}{4}} 
\newcommand{\ut}{{\tilde u}}        
\newcommand{\vac}{\ket{0}}          
\newcommand{\vev}[1]{\braCket{0}{#1}{0}} 
\newcommand{\vf}{\varphi}           
\newcommand{\W}{\mathbb{W}}        
\newcommand{\x}{\times}             
\newcommand{\7}{\dagger}            
\def\wick:#1:{\mathopen:#1\mathclose:} 
\def\<#1,#2>{\langle#1,#2\rangle}   

\theoremstyle{plain}

\theoremstyle{definition}

\begin{document}

\title{Lectures on BRS invariance \\ for massive boson fields}

\author{
Jos\'e M. Gracia-Bond\'{\i}a\,\dag\
\\
\dag Departamento de F\'{\i}sica Te\'orica,
\\
Universidad de Zaragoza, Zaragoza 50009, Spain
}

\date{\today}

\maketitle

\centerline{\date}

\begin{abstract}
These notes correspond to lectures given at the Villa de Leyva Summer
School in Colombia (July 2007).  Our main purpose in this short course
on BRS invariance of gauge theories is to illuminate corners of the
theory left in the shade by standard treatments.  The plan is as
follows.  First a review of Utiyama's ``general gauge theory''.
Promptly we find a counterexample to it in the shape of the massive
spin-1 St\"uckelberg gauge field.  This is not fancy, as the massive
case is the most natural one to introduce BRS invariance in the
context of free quantum fields.  Mathematically speaking, the first
part of the course uses Utiyama's notation, and thus has the flavour
and non-intrinsic notation of standard physics textbooks.  Next we
deal with boson fields on Fock space and BRS invariance in connection
with the existence of Krein operators; the attending rigour points are
then addressed.
\end{abstract}

\tableofcontents

\section{Utiyama's method in classical gauge theory}

\subsection{A historical note}

Ryoyu Utiyama developed non-abelian gauge theory early in 1954 in
Japan, almost at the same time that Yang and Mills~\cite{YM} did at
the Princeton's Institute for Advanced Study (IAS), that Utiyama was
to visit later in the year.  Unfortunately, Utiyama chose not to
publish immediately, and upon his arrival at IAS on September of that
year, he was greatly discouraged to find he had apparently just been
``scooped''.

In fact, he had not, or not entirely.  He writes: ``(In March 1955), I
decided to return to the general gauge theory, and took a closer look
at Yang's paper, which had been published in 1954.  At this moment I
realized for the first time that there was a significant difference
between Yang's theory and mine.  The difference was that Yang had
merely found an example of non-abelian gauge theory whereas I had
developed a general idea of gauge theory that would contain gravity as
well of electromagnetic theory.  Then I decided to publish my work by
translating it into English, and adding an extra section where Yang's
theory is discussed as an example of my general
theory''~\cite{OneRisingSun}.

Utiyama's article appeared on the March~1, 1956 issue of the Physical
Review~\cite{MeanwhileInJapan}, and is also is reprinted in the book
by the late Lochlainn O'Raifeartaigh~\cite{OneRisingSun}, where the
foregoing (and other) interesting historical remarks are made.

As Utiyama himself does above, most people who read his paper focused
on the kinship there shown between gravity and gauge theory.  This is
in some sense a pity, because in contrast with ``textbook'' treatments
of Yang--Mills theories ---see~\cite{Pesky} for just one example---
which manage to leave, despite disguises of relatively sophisticated
language, a strong impression of arbitrariness, Utiyama strenously
tried to \textit{derive} gauge theory from first principles.  The most
important trait of~\cite{MeanwhileInJapan} is that he asks the right
questions from the outset, as to what happens when a Lagrangian
invariant with respect to a global Lie group $G$ is required to become
invariant with respect to the local group $G(x)$.  What kind of new
(gauge) fields need be introduced to `maintain' the symmetry?  What is
the form of the new Lagrangian, including the interaction?  His answer
is that the gauge field \textit{must} be a spacetime vector field on
which $G(x)$ acts by the adjoint representation, transforming in such
a way that a covariant derivative exists.  To our knowledge, the
Utiyama argument is reproduced only in a couple of modern texts; such
are~\cite{Frampton} and~\cite{Trickyguys}.  I have profited from the
excellent notes~\cite{MeanwhileInBaturria} as well.

One can speculate that, if the sequence of events had been slightly
different, more attention would have been devoted to the theoretical
underpinnings of the accepted dogma. It is revealing, and another
pity, that Utiyama's later book in Japanese on the general gauge
theory has never been translated.

\subsection{The Utiyama analysis, first part}

The starting point for Utiyama's analysis is a Lagrangian
$$
\L(\vf_k,\del_\mu\vf_k),
$$
depending on a multiplet of fields $\vf_k$ and their first
derivatives, \textit{globally} invariant under a group~$G$ (of ``gauge
transformations of the first class'') with $n$ independent
parametres~$\th^a$.  The group is supposed to be compact.  We denote
by $f^{abc}$ the structure constants of its Lie algebra $\g$; that is
$\g$ possesses generators~$T^a$ with commutation relations
$$
[T^a, T^b] = f^{abc}T^c, \sepword{with} f^{abc} = -f^{bca},
$$
and the Jacobi identity:
\begin{equation}
f^{abd}f^{dce} + f^{bcd}f^{dae} + f^{cad}f^{dbe} = 0
\label{eq:dying-ember}
\end{equation}
holds.  We assume that the $T^a$ can be chosen in such a way that
$f^{abc}$ is antisymmetric in all the three indices.  This means that
the adjoint representation of~$\g$ is semisimple, that is, $\g$ is
reductive~\cite[Chapter~15]{VacuumCleanerBis}.  Close by the identity,
an element $g\in G$ is of the form $\exp(T^a\th^a)$.

The invariance is to be extended to a group~$G(x)$ ---of ``gauge
transformations of the second class''--- depending on local
parametres~$\th^a(x)$, in such a way that a new Lagrangian
$\L(\vf_k,\del_\mu\vf_k,A)$ invariant under the wider class of
transformations is uniquely determined.  Utiyama's questions are:

\begin{itemize}
\item
What new field $A(x)$ needs to be introduced?

\item
How does $A(x)$ transforms under $G(x)$?

\item
What are the form of the interaction and the new Lagrangian?

\item
What are the allowed field equations for~$A(x)$?
\end{itemize}

The global invariance is given to us under the form:
\begin{equation}
\dl\vf_k(x) = T^a_{kl}\vf_l(x)\th^a; \sepword{now we want to consider}
\dl\vf_k(x) = T^a_{kl}\vf_l(x)\th^a(x),
\label{eq:initial-condition}
\end{equation}
for~$1\le a\le n$. This last transformation in general does not leave
$\L$ invariant. Let us first learn about the constraints imposed on
the Lagrangian density by the assumed global invariance. One has
\begin{equation}
0 = \dl\L = \frac{\del\L}{\del \vf_k}\,\dl\vf_k +
\frac{\del\L}{\del(\del_\mu\vf_k)}\,\dl\,\del_\mu\vf_k,
\label{eq:no-has-visto-nada}
\end{equation}
where now
\begin{equation}
\dl\,\del_\mu\vf_k = \del_\mu\,\dl\vf_k =
T^a_{kl}\del_\mu\vf_l(x)\th^a(x) + T^a_{kl}\vf_l(x)\del_\mu\th^a(x).
\label{eq:weak-link}
\end{equation}
With a glance back to~\eqref{eq:no-has-visto-nada}
and~\eqref{eq:weak-link}, we see that
\begin{equation}
\dl\L =
\frac{\del\L}{\del(\del_\mu\vf_k)}T^a_{kl}\vf_l(x)\del_\mu\th^a(x)
\ne0.
\label{eq:to-be-or-not-to-be}
\end{equation}

Then it is necessary to add new fields $A'_p,p=1,\dots,M$ in the
Lagrangian, a process which we write as
$$
\L(\vf_{k},\del_\mu\vf_k) \longrightarrow
\L'(\vf_{k},\del_\mu\vf_k,A'_p).
$$
The question is, how do the new fields transform?  We assume not only
a term of the form~\eqref{eq:weak-link} but also a derivative term
in~$\th^a(x)$ ---indeed the latter will be needed to compensate the
right hand side of~\eqref{eq:to-be-or-not-to-be}:
\begin{equation}
\dl A'_p = U^a_{pq}A'_q\th^a + C^{a\mu}_p\del_\mu\th^a.
\label{eq:madre-de-luna}
\end{equation}
Here $C^{a\mu}_p$ and the $U^a_{pq}$ are constant matrices, for the
moment unknown.  The requirement~is
\begin{equation*}
0 = \dl\L' = \frac{\del\L'}{\del \vf_k}\,\dl\vf_k +
\frac{\del\L'}{\del(\del_\mu\vf_k)}\,\del_\mu\dl\vf_k +
\frac{\del\L'}{\del A'_p}\,\dl A'_p,
\end{equation*}
boiling down to
\begin{align}
\dl\L' &= \biggl[\frac{\del\L'}{\del\vf_k}T^a_{kl}\vf_l +
\frac{\del\L'}{\del(\del_\mu\vf_k)}T^a_{kl}\del_\mu\vf_l +
\frac{\del\L'}{\del A'_p}U^a_{pq}A'_q\biggr]\th^a
\nonumber\\
&+ \biggl[\frac{\del\L'}{\del(\del_\mu\vf_k)}T^a_{kl}\vf_l +
\frac{\del\L'}{\del A'_p}C^{a\mu}_p\biggr]\del_\mu\th^a = 0.
\label{eq:bittersweet}
\end{align}
The coefficients must vanish separately, as the $\th^a$ an their
derivatives are arbitrary.  The coefficient of~$\del_\mu\th^a$
gives~$4n$ equations involving $A'_p$, and hence to determine the~$A'$
dependence uniquely one needs $M=4n$ components.  Furthermore, the
matrix $C^{a\mu}_p$ must be nonsingular.  We have then an inverse:
$$
C^{a\mu}_p{C^{-1}}^a_{\mu q} = \dl_{pq}; \qquad
{C^{-1}}^a_{\mu p}C_p^{b\nu} = \dl^\nu_\mu\dl^{ab}.
$$

Define the gauge (potential) field
\begin{equation}
A^a_\mu = \frac{1}{g}{C^{-1}}^a_{\mu p}\,A'_p, \sepword{with
inverse} A'_p = gC^{a\mu}_p A^a_\mu.
\label{eq:claro-de-luna}
\end{equation}
Before proceeding, note that~\eqref{eq:madre-de-luna}
and~\eqref{eq:claro-de-luna} together imply
$$
\dl A^a_\mu = \big({C^{-1}}^a_{\mu p}U^c_{pq}C^{b\nu}_q\big)
A^b_\nu\th^c + \frac{\del_\mu\th^a}{g} =: (S^a_\mu)^{cb\nu}
A^b_\nu\th^c + \frac{\del_\mu\th^a}{g}.
$$
Clearly from~\eqref{eq:bittersweet} we have
$$
\frac{\del\L'}{\del(\del_\mu\vf_k)}T^a_{kl}\vf_l +
\frac{1}{g}\frac{\del\L'}{\del A_\mu^a} = 0.
$$
Hence only the combination (called the covariant derivative)
$$
D_\mu\vf_k := \del_\mu\vf_k - gT^a_{kl}\vf_l A^a_\mu
$$
occurs in~$\L'(\vf_{k},\del_\mu\vf_k,A'_p)$, and we rewrite:
$$
\L'(\vf_k,\del_\mu\vf_k,A'_p) \longrightarrow
\L''(\vf_k,D_\mu\vf_k).
$$
Moreover, it follows
\begin{align*}
\frac{\del\L'}{\del\vf_k} &= \frac{\del\L''}{\del\vf_k} -
g\frac{\del\L''}{\del(D_\mu\vf_l)}T^a_{lk}A^a_\mu;
\\
\frac{\del\L'}{\del(\del_\mu\vf_k)} &=
\frac{\del\L''}{\del(D_\mu\vf_k)};
\\
\frac{\del\L'}{\del A'_p} &=
-\frac{\del\L''}{\del(D_\mu\vf_k)}T^a_{kl}\vf_l{C^{-1}}^a_{\mu p}.
\end{align*}

Now we look at the vanishing coefficient of~$\th^a$ occurring
in~$\dl\L'$ in~\eqref{eq:bittersweet}. By use of the last set of
equations:
\begin{align}
0 &= \frac{\del\L''}{\del\vf_k}T^a_{kl}\vf_l -
g\frac{\del\L''}{\del(D_\mu\vf_m)}T^b_{mk}T^a_{kl}A^b_\mu\vf_l
\nonumber \\
&+ \frac{\del\L''}{\del D_\mu\vf_k}T^a_{kl}\del_\mu\vf_l -
g\frac{\del\L''}{\del(D_\mu\vf_m)}T^c_{ml}\vf_l{C^{-1}}^c_{\mu p}
U^a_{pq}C^{b\nu}_q A^b_\nu
\nonumber \\
&= \frac{\del\L''}{\del\vf_k}T^a_{kl}\vf_l + \frac{\del\L''}{\del 
D_\mu\vf_k}T^a_{kl}D_\mu\vf_l 
\nonumber \\
&- g\frac{\del\L''}{\del(D_\mu\vf_m)}\big[T^b_{mk}T^a_{kl}A^b_\mu\vf_l
- T^a_{mk}T^b_{kl}A^b_\mu\vf_l + T^c_{ml}(S^c_\mu)^{ab\nu}A^b_\nu\vf_l
\big].
\label{eq:breakfast-at-Tiffany's}
\end{align}

We are come thus to the crucial (and delicate) point.  It seems that
the two first terms in~\eqref{eq:breakfast-at-Tiffany's} cancel each
other by global invariance (!)  if we identify
$$
\L''(\vf_k,D_\mu\vf_k) = \L(\vf_k,D_\mu\vf_k). 
$$
Utiyama~\cite{MeanwhileInJapan} writes here: ``This particular choice
of $\L''$ is due to the requirement that when the field~$A$ is assumed
to vanish, we must have the original Lagrangian''.  It seems to me,
however, that covariance of $D_\mu\vf_k$ is implicitly required.  The
whole procedure is at least consistent: the vanishing of the last term
in~\eqref{eq:breakfast-at-Tiffany's} allows us to identify
$$
(S^c_\mu)^{ab\nu} = f^{abc}\dl_\mu^\nu.
$$
This implies in the end
\begin{equation}
\dl A^a_\mu = f^{cba}A_\mu^b\th^c + \frac{\del_\mu\th^a}{g}.
\label{eq:bottom-line}
\end{equation}
As a consequence we obtain that $D_\mu\vf_k$ indeed is a covariant
quantity, in the sense of~\eqref{eq:weak-link}:
\begin{align*}
\dl(D_\mu\vf_k) &= \dl(\del_\mu\vf_k - gT^a_{kl}A^a_\mu\vf_l) =
\del_\mu(T^a_{kl}\th^a\vf_l) - gf^{cba}T^a_{km}A_\mu^b\th^c\vf_m
\\
&- T^a_{kl}\del_\mu\th^a\vf_l - gT^b_{kl}T^c_{lm}A^b_\mu\th^c\vf_m =
T^a_{kl}\th^a\del_\mu\vf_l - gT^c_{kl}T^b_{lm}A^b_\mu\th^c\vf_m
\\
&= T^a_{kl}\th^a(D_\mu\vf_l).
\end{align*}
(In summary, Utiyama's argument here looks a bit circular to us; but
all is well in the~end.)

\subsection{Final touches to the Lagrangian}

The local Lagrangian of the matter fields contains in the bargain the
interaction Lagrangian between matter and gauge fields.  The missing
piece is the Lagrangian for the ``free'' $A$-field.  Next we
investigate its possible type.  Call the sought for Lagrangian
$\L_0(A^a_{\nu}, \del_{\mu}A^a_{\nu})$.  The invariance (under the 
local group of internal symmetry) postulate
together with~\eqref{eq:bottom-line} in detail says:
\begin{align*}
0 &= \biggl[\frac{\del\L_0}{\del A^a_{\nu}}
f^{cba}A^b_\nu + \frac{\del\L_0}{\del(\del_\mu A^a_\nu)}
f^{cba}\del_\mu A^b_\nu\biggr]\th^c
\\
&\qquad +\biggl[\frac{\del\L_0}{\del(\del_{\mu}A^a_\nu)}
f^{cba}A^b_\nu + \frac{1}{g}\,\frac{\del\L_0}{\del A^c_{\mu}}\biggr]
\del_\mu\th^c
\\
&\qquad +\frac{1}{g}\frac{\del\L_0}{\del(\del_{\mu}A^c_{\nu})}
\,\del_{\mu\nu}\th^c.
\end{align*}
As the $\th^c$ are arbitrary again, one concludes that
\begin{align}
\frac{\del\L_0}{\del A^a_{\nu}}f^{cba}A^b_\nu +
\frac{\del\L_0}{\del(\del_\mu A^a_\nu)}f^{cba}\del_\mu A^b_\nu &= 0,
\label{eq:mama-de-TarzanI}
\\
\frac{\del\L_0}{\del(\del_{\mu}A^a_{\nu})}f^{cba}A^b_\nu +
\frac{1}{g}\,\frac{\del\L_0}{\del A^c_{\mu}} &= 0,
\label{eq:mama-de-TarzanII}
\\
\frac{\del\L_0}{\del(\del_{\mu}A^a_{\nu})} +
\frac{\del\L_0}{\del(\del_{\nu}A^a_{\mu})} &= 0.
\label{eq:mama-de-TarzanIII}
\end{align}
Introduce provisionally:
$$
\A^a_{\mu\nu} := \del_{\mu}A^a_{\nu} - \del_{\nu}A^a_{\mu}.
$$
Then~\eqref{eq:mama-de-TarzanII} is rewritten
$$
\frac{\del\L_0}{\del A^c_{\mu}} +
2g\frac{\del\L}{\del(\A^a_{\mu\nu})}f^{cba}A^b_\nu=0.
$$
It ensues that the only combination occurring in the Lagrangian is
\begin{equation}
F^c_{\mu\nu} := \A^c_{\mu\nu} - \thalf gf^{abc}(A^a_{\mu}A^b_{\nu} -
A^a_{\nu}A^b_{\mu}).
\label{eq:estas-avisado}
\end{equation}
One may write then
$$
\L_0(A^a_\nu,\del_\mu A^a_\nu) = \L'_0(F^a_{\mu\nu}).
$$
Parenthetically we note
\begin{equation*}
F^a_{\mu\nu} + F^a_{\nu\mu} = 0.
\end{equation*}

Now,
$$
\frac{\del\L_0}{\del(\del_\mu A^a_\nu)} = 2\frac{\del\L'_0}{\del
F^a_{\mu\nu}}; \qquad \frac{\del\L_0}{\del A^b_\mu} =
2\frac{\del\L'_0}{\del F^c_{\mu\nu}}f^{abc}A^a_\nu.
$$
Thus, by use of~\eqref{eq:dying-ember},
formula~\eqref{eq:mama-de-TarzanI} means
\begin{equation}
\frac{\del\L'_0}{\del F^c_{\mu\nu}}f^{abc}F^a_{\mu\nu} = 0,
\label{eq:great-bis}
\end{equation}
for $1\le b\le n$.  This is left as an exercise.  Also, by use of the
identity of Jacobi again, one obtains
\begin{equation}
\dl F^c_{\mu\nu} = f^{abc}F^b_{\mu\nu}\th^a.
\label{eq:dirty-trick}
\end{equation}
This is a covariance equation similar to~\eqref{eq:weak-link}; its
proof is an exercise as well.

Equation~\eqref{eq:great-bis} is as far as we can go with the general
argument. The simplest Lagrangian satisfying this condition is the
quadratic in~$F^a_{\mu\nu}$ one:
\begin{equation}
\L_{\rm YM} := -\tquarter F^a_{\mu\nu}F^{a\,\mu\nu} \sepword{implying}
F^a_{\mu\nu} = -\frac{\del\L_{\rm YM}}{\del(\del_{\mu}A^a_{\nu})}.
\label{eq:hacia-atras-sin-ira}
\end{equation}
The last equation is consistent with~\eqref{eq:mama-de-TarzanIII}.
Note that $\dl\L_{\rm YM}=0$ from~\eqref{eq:dirty-trick} is obvious.

If now we define
\begin{equation}
J^{c\mu} = g f^{abc}\frac{\del\L_{\rm
YM}}{\del(\del_{\mu}A^a_{\nu})}A^b_\nu,
\label{eq:current-famous}
\end{equation}
then from~\eqref{eq:mama-de-TarzanI} again:
\begin{equation}
\del_\mu J^{a\mu} = 0;
\label{eq:great-three}
\end{equation}
and from~\eqref{eq:mama-de-TarzanII}:
\begin{equation}
\del^\nu F^a_{\mu\nu} = J^a_\mu,
\label{eq:great-two}
\end{equation}
by use of the equations of motion in both cases.

Let us take stock of what we obtained. 

\begin{itemize}
\item
Formula~\eqref{eq:current-famous} tells us that (in this non-nabelian
case) a self-interaction current~$J_\mu$ exists, and gives us an
explicit expression for it.

\item
Equation~\eqref{eq:great-three} furthermore shows that the current is
conserved.  Such a conservation equation, involving ordinary
derivatives instead of covariant ones, does not look very natural
perhaps, and is not so easy to prove directly ---see the discussion
in~\cite[Section~12-1-2]{IZ}.  This is the content of Noether's second
theorem as applied in the present context.

\item
We observe that~\eqref{eq:great-two} is the field equation in the
absence of matter fields.
\end{itemize}

The full Lagrangian is $\L(\vf_k,D_\mu\vf_k)+\L'_{\rm YM}$.  One can
proceed now to verify the invariance of it under the local
transformation group and study the corresponding conserved currents.
It should be clear that the conserved currents arising from local
gauge invariance are exactly those following from global gauge
invariance.  Left as exercise.

\subsection{The electromagnetic field}

We illustrate only with the simplest example, as our main purpose is
to produce a `counterexample' pretty soon. Let a Dirac spinor field of
mass~$M$ be given:
$$
\L = \tihalf[\owl\psi\ga^\mu\del_\mu\psi -
\del_\mu\owl\psi\,\ga^\mu\psi] - \owl\psi M\psi.
$$
(Borrowing the frequent notation $A\overleftrightarrow {\del^\a}B =
A\del^\a B - (\del^\a A)B$, one can write this as well as
$$
\tihalf\owl\psi\overleftrightarrow{\del_\mu}\ga^\mu\psi - \owl\psi
M\psi.)
$$
This is invariant under the global abelian group of phase
transformations
$$
\owl\psi(x) \mapsto e^{i\th}\psi(x);
\quad \psi(x) \mapsto e^{-i\th}\psi(x);
$$
or, infinitesimally,
$$
\dl\owl\psi = i\owl\psi\th; \quad \dl\psi = -i\psi\th.
$$
This leads to the covariant derivatives
$$
D_\mu\owl\psi = \del_\mu\owl\psi - igA_\mu\owl\psi; \quad
D_\mu\psi = \del_\mu\psi + igA_\mu\psi.
$$
In conclusion, the original Lagrangian gets an interaction piece
$-g\owl\psi\ga^\mu A_\mu\psi$; with invariance of the new Lagrangian
thanks to $\dl A_\mu = \del_\mu\th/g$.  The full locally invariant
Lagrangian is
$$
\tihalf[\owl\psi\ga^\mu\del_\mu\psi - \del_\mu\owl\psi\ga^\mu\psi] -
g\owl\psi\ga^\mu A_\mu\psi - \owl\psi M\psi - \tquarter
F_{\mu\nu}F^{\mu\nu}.
$$
One can find now the associated electromagnetic current.  This is the
last exercise of this section.

\subsection{The original Yang-Mills field}

Consider an isospin doublet of spinor fields:
$$
\psi = (\psi_k) = \begin{pmatrix} \psi_1 \\ \psi_2 \end{pmatrix},
$$
with free Lagrangian
$$
\tihalf[\owl\psi_k\ga^\mu\del_\mu\psi_k -
\del_\mu\owl\psi_k\ga^\mu\psi_k] - \owl\psi_k M\psi_k.
$$
This is invariant under the global $SU(2)$ group; with $\sigma^a$
denoting as usual the Pauli matrices:
$$
\psi_k \mapsto e^{-ig\th^a\sigma^a/2}\big|_{kl}\psi_l; \qquad
\owl\psi_k \mapsto \owl\psi_l\,e^{ig\th^a\sigma^a/2}\big|_{lk}.
$$
Infinitesimally,
$$
\dl\psi_k = T^a_{kl}\psi_l\th^a, \sepword{with} T^a_{kl} =
-\frac{ig}{2}\sigma^a_{kl}.
$$
We have $f^{abc}=g\eps^{abc}$ for this group. The Lagrangian becomes 
gauge invariant through the replacement
$$
\del_\mu\psi_k \mapsto D_\mu\psi_k = \del_\mu\psi_k +
\frac{ig}{2}\sigma^a_{kl}\psi_l A^a_\mu;
$$
That is, the triplet of vector fields is the gauge (potential) field,
the number of gauge field components being equal to the number of
symmetry generators.  Note the slight difference in the introduction
of the coupling constant of the gauge field with the spinor field and
itself.

The full locally invariant Lagrangian is
\begin{equation*}
\tihalf[\owl\psi_k\ga^\mu\del_\mu\psi_k -
\del_\mu\owl\psi_k\ga^\mu\psi_k] - \owl\psi_k M\psi_k -\tquarter
F^a_{\mu\nu}F^{a\,\mu\nu} - \frac{g}{2} \owl\psi_k\ga^\mu\sigma^a_{kl}
\psi_l A^a_\mu,
\end{equation*}
with $F^a_{\mu\nu}$ given by~\eqref{eq:estas-avisado}.  The current
\begin{align*}
J_\mu^a &= - \frac{g}{2} \owl\psi_k\ga^\mu\sigma^a_{kl} \psi_l -
g\eps^{abc}A^c_\nu\big[\del_\mu A^b_\nu - \del_\nu A^b_\mu -
\frac{g}{2}\eps^{bde}(A^d_\mu A^e_\nu - A^d_\nu A^e_\mu)\big]
\\
&= - \frac{g}{2} \owl\psi_k\ga^\mu\sigma^a_{kl} \psi_l -
g\eps^{abc}A^c_\nu\big(\del_\mu A^b_\nu - \del_\nu A^b_\mu) + 
g^2(A^a_\mu(AA) + A^c_\mu A^c_\nu A^a_\nu),
\end{align*}
with $AA:=A^c_\nu\,A^{c\nu}$, is conserved.

\section{Massive vector fields}

\subsection{What is wrong with the Proca field?}

The starting point in relativistic quantum physics is Wigner's theory
of particles~\cite{WignerSource} as positive-energy irreps of the
Poincar\'e group with finite spin/helicity.  The transition to local
free fields is made through intertwiners between the Wigner
representation matrices and the matrices of covariant Lorentz group
representations.  Therefore, following standard
notations~\cite{VacuumCleaner}, the general form of a quantum field is
\begin{align*}
\vf_l(x) &= \vf_l^{(-)}(x) + \vf_l^{(+)}(x) \sepword{with}
\\
\vf_l^{(-)}(x) &= (2\pi)^{-3/2}\sum_{\sigma,n}\int
d\mu_m(k)\,u_l(k,\sigma,n)e^{-ikx}a(k,\sigma,n);
\\
\vf_l^{(+)}(x) &= (2\pi)^{-3/2}\sum_{\sigma,n}\int
d\mu_m(k)\,v_l(k,\sigma,n)e^{ikx}a^\7(k,\sigma,n);
\end{align*}
with $d\mu_m(k)$ the usual Lorentz-invariant measure on the mass~$m$
hyperboloid in momentum space and~$n$ standing for particle species.
Leaving the latter aside, the other labels are of
representation-theoretic nature.  Operator solutions to the wave
equations carry the following labels, in all: the Poincar\'e
representation $(m,s)$, that gives the the mass shell condition and
the spin~$s$; the $(k,\sigma)$, with the range of $\sigma$ determined
by~$s$, label the momentum basis states; the $(u,v)$ are Lorentz
representation labels, usually appearing as a superscript indicating
the tensorial or spinorial character of that solution.  The $c$-number
functions~$u_l,v_l$ in the plane-wave expansion formulae are the
coefficient functions or intertwiners, connecting the set of creation
or absorption operators $a^{\#}(k,\sigma)$, transforming as the
irreducible representation $(m,s)$ of the Poincar\'e group, to the set
of field operators $\vf_l(x)$, transforming as a certain
finite-dimensional ---thus nonunitary--- irrep of the Lorentz group.
We have thus in the vector field case
\begin{align*}
\vf^{(-)\mu}(x) &= (2\pi)^{-3/2}\sum_{\sigma}\int
d\mu_m(k)\,u^\mu(k,\sigma)e^{-ikx}a(k,\sigma);
\\
\vf^{(+)\mu}(x) &= (2\pi)^{-3/2}\sum_{\sigma}\int
d\mu_m(k)\,v^\mu(k,\sigma)e^{ikx}a^\7(k,\sigma).
\end{align*}
We neglect to consider in the notation any colour quantum number for a
while.

For the spin of the particle described by the vector field both the
values~$j=0$ and~$j=1$ are possible. In the first case, at~$\vec k=0$
only $u^0,v^0$ are non-zero, and, dropping the label~$\sigma$, we have
by Lorentz invariance
$$
u^\mu(k) \propto ik^\mu; \quad v^\mu(k) \propto - ik^\mu,
$$
and therefore $\vf^\mu(x)=\del^\mu\vf(x)$ for some scalar field~$\vf$.
In the second case, only the space components $u^j,v^j$ are not
vanishing at~$\vec k=0$, and we are led to
\begin{equation}
\vf^{(-)\mu}(x) = {\vf^{(+)\mu}}^\7(x) =
(2\pi)^{-3/2}\sum_{\sigma=1}^3
\int d\mu_m(k)\,\eps^\mu(k,\sigma)e^{-ikx}a(k,\sigma),
\label{eq:in-articulo-mortis}
\end{equation}
with $\eps^\mu$ suitable (spacelike, normalized, orthogonal
to~$k_\mu$, also real) polarization vectors, so that
\begin{equation}
\sum_{\sigma=1}^{3}\eps_\mu(k,\sigma)\eps_\nu(k,\sigma) =
-g_{\mu\nu}+\frac{k_\mu k_\nu}{m^2}.
\label{eq:no-trigo-limpio}
\end{equation}
On the right hand side we have the projection matrix on the space
orthogonal to the four vector $k^\mu$. This may be rewritten
$$
\sum_{\sigma=0}^{\sigma=3}g_{\sigma\sigma}\eps_\mu(k,\sigma)
\eps_\nu(k,\sigma) = g_{\mu\nu},
$$
with the definition~$\eps_\mu(k,0)=k_\mu/m$. With this treatment,
we have the equations
\begin{equation*}
(\square + m^2)\vf^\mu(x) = 0; \qquad \del_\mu\vf^\mu(x) = 0.
\end{equation*}
The last one ensures that one of the four degres of freedom
in~$\vf^\mu$ is elimimated. However,
eventually~\eqref{eq:no-trigo-limpio} leads to the commutation
relations for the Proca field of the form
$$
[\vf^\mu(x), \vf^\nu(y)] =
i\biggl(g^{\mu\nu}+\frac{\del^\mu\del^\nu}{m^2}\biggr)D(x-y).
$$
In momentum space this is constant as $|k|\uparrow\infty$, which bodes
badly for renormalizability. The Feynman propagator is proportional to
$$
\frac{g_{\mu\nu} - k_\mu k_\nu/m^2}{k^2-m^2};
$$
there is moreover a troublesome extra term, that we leave aside.

The argument for non-renormalizability is as follows.  Suppose that,
as in the exampls of the previous section, the vector field is coupled
with a conserved current made out of spinor fields.  Consider an
arbitrary Feynman graph with $E_F$ external fermion lines, $I_F$
internal ones, and respectively $E_B,I_B$ boson lines.  The assumption
says two fermion lines and one boson line meet at each vertex.  The
number of vertices is thus
$$
V = 2I_B + E_B = \thalf(2I_F + E_F).
$$
Since there is a delta function for each vertex, one of them
corresponding to overall momentum conservation, and each internal line
has an integration over its moment, by eliminating $I_F,I_B$ the
superficial degree of divergence is
$$
D = -4(V - 1) + 3I_F + 4I_B = 4 + V - 3E_F/2 - 2E_B.
$$
This shows that, no matter how many external lines are, the degree of
divergence can be made arbitrarily large.

The difficulty is with the intertwiners, whose dimension does not
allow to usual renormalizability condition.  The idea is then to cure
this by a cohomological extension of the Wigner representation space
for massive spin~1 particles.  This involves both the St\"uckelberg
field and the ghost fields, already at the level of the description of
free fields.  The nilpotency condition $s^2=0$ for the BRS
operator~$s$ will yield a cohomological representation for the
physical Hilbert space $\ker s/\ran s$, which, as we shall see later,
is the (closure of) the space of transversal vector wavefunctions.  On
that extended Hilbert space the renormalizability problem fades away.
This goes in hand with a philosophy of primacy of a quantum character
for the gauge principle, that should be read backwards into classical
field theory; fibre bundle theory is no doubt elegant, but not
intrinsic from this viewpoint.  (For massless particles, the
situtation is worse in that problematic aspects of the use of vector
potentials in the local description of spin~1 particles show up
already in the covariance properties of photons and gluons.)

\subsection{What escaped through the net}

Another unsung hero of quantum field theory is the Swiss physicist
Ernst Carl Gerlach St\"uckelberg, bar\'on von Breidenbach.  He found
himself among the pioneers of the `new' Quantum Mechanics; at the end
of the twenties, while working in Princeton with Morse, he was the one
to explain the continuous spectrum of molecular hydrogen.  At his
return to Europe in 1933, he met Wentzel and Pauli for the first time.
St\"uckelberg stayed in Zurich for two years before accepting a
position at Gen\`eve.  He turned to particle physics, where he will
among other things contribute, according to his
obituary~\cite{Rivier}, the meson hypothesis (unpublished at the time
because of Pauli's criticism, and usually associated with Yukawa), the
causal propagator (better known as the Feynman propagator) and the
renormalization group~\cite{HPA1,HPA2}.  Also by St\"uckelberg, not
underlined in~\cite{Rivier}, are the first formulation of baryon
number conservation; the first sketch of what is called nowadays
Epstein--Glaser renormalization~\cite{EGAvantLaLettre} ---towards
which, according to the account in~\cite{BlueBook}, Pauli was better
disposed--- and the \textit{St\"uckelberg field}~\cite{OldUnknown},
which concerns us here.

\smallskip

We have seen the extreme care that Utiyama put in deriving the precise
form of gauge theory as a theorem.  However, already at the moment
that he published it, his result was false.  That something that
escapes through Utiyama's net is St\"uckelberg's gauge theory for
massive spin~1 particles.

In the old paper~\cite{Grognard} Pauli rather dismissively had given a
short account of that before plunging into the Proca field; although
anyone who has tried to work with the latter rapidly realizes it is
good for nothing.  There are several natural ways to discover the
St\"uckelberg gauge field, even after one has been miseducated by
textbooks ---like~\cite{VacuumCleaner}--- into exclusively learning
about the Proca field.  A principled quantum approach is contained
\textit{in embrio} in the paper~\cite{CabezondelaSal}, where the
starting point is Wigner's picture of the unitary irreps of the
Poincar\'e group.  In the book by Itzykson and Zuber, the
St\"uckelberg method is used time and
again~\cite[pp.~136,~172,~610]{IZ} to smooth the $m\downarrow0$ limit
and exorcise infrared troubles.  A very useful reference for the
St\"uckelberg field is the review~\cite{Altabonazo}.  We have been
inspired also by~\cite{Steklovazo}.

\subsection{The St\"uckelberg field and Utiyama's test}

Actually, there is no logical fault in the Lagrangian approach by
Utiyama.  Where he goes astray is only in the ``initial
condition''~\eqref{eq:initial-condition}.  We next try to find the
St\"uckelberg field by the Utiyama path; that is, whether we actually
could have derived the existence of the field~$B$ using the arguments
of subsection~1.2.  We do this for an abelian theory.  Assume that a
globally $G\equiv U(1)$-invariant model of a Dirac fermion of mass~$M$
and a real vector field of mass~$m$ are given:
\begin{align*}
\L_0 &= \frac{i}{2}(\owl\psi\ga^\mu\del_\mu\psi -
\del_\mu\owl\psi\,\ga^\mu\psi) - \owl\psi M\psi + \thalf m^2 A_\mu
A^\mu + \L_{\rm kin}(\del_\nu A_\mu)
\\
&=: \L_{0,{\rm f}} + \L_{0,{\rm phmass}} + \L_{\rm kin},
\end{align*}
with an obvious notation.  This is obviously a model for
(non-interacting) massive photon electrodynamics.  Here $\L_{\rm kin}$
is the kinetic energy term for the photon, of the
form~\eqref{eq:hacia-atras-sin-ira}.  This Lagrangian is invariant
under the global gauge transformations:
$$
A_\mu(x) \mapsto A_\mu(x); \quad \owl\psi(x) \mapsto e^{i\th}\psi(x);
\quad \psi(x) \mapsto e^{-i\th}\psi(x);
$$
or, infinitesimally,
$$
\dl A_\mu = 0; \quad \dl\owl\psi = i\owl\psi\th; \quad \dl\psi =
-i\psi\th.
$$

Now the Utiyama questions come in: what new (gauge) fields need be
introduced?  How do they transform under $G(x)$?  What is the form of
the interaction, and what is the new Lagrangian?  To save spacetime,
we restart from
\begin{align*}
&\tihalf[\owl\psi\ga^\mu\del_\mu\psi - \del_\mu\owl\psi\ga^\mu\psi] -
\owl\psi\ga^\mu A_\mu\psi - \owl\psi M\psi + \thalf m^2 A_\mu A^\mu 
\\
&- \tquarter (\del_\mu A_\nu -\del_\nu A_\mu)(\del^\mu A^\nu -\del^\nu
A^\mu) =: \L_{\rm f} + \L_{0,{\rm phmass}} + \L_{\rm kin}.
\end{align*}
The multiplet of fields includes now
\begin{equation}
\vf = \begin{pmatrix} \owl\psi \\ \psi \\ A^\mu \end{pmatrix}
\sepword{transforming as} \dl\vf = \begin{pmatrix} i\owl\psi \th(x) \\
-i\psi \th(x) \\ \del^\mu\th(x) \\ \end{pmatrix};
\label{eq:talon-de-Aquiles}
\end{equation}
where of course we required a variation of the QED type for
the~$A_\mu$.  For simplicity we have put $g=1$.  However, still
$$
\dl\L_0 = \frac{\del\L_{0,{\rm phmass}}}{\del A_\mu}\,\dl A_\mu =
m\del_\mu\th \ne 0.
$$

It seems that, when vector fields are conjured \textit{ab~initio},
further infinitesimal gauge transformations of the form
\begin{equation}
\dl\vf_k = \A_{kc}\th_c + \B^\nu_{kc}\del_\nu\th_c,
\label{eq:untrodden-path}
\end{equation}
need to be considered.  Here we have a particular case, with a trivial
colour index~$c$; with $\vf_k\to A_\mu;\;\A_{\mu}$ vanishing; and
$\B^\nu_\mu=\dl^\nu_\mu$.

There is no need to involve other parts of the Lagrangian than
$\L_{0,{\rm phmass}}$ in the remaining calculation.  We need an extra
vector field.  It is natural to think that it be fabricated from the
derivatives of a scalar~$B$, and we write:
$$
\L_{0,{\rm phmass}}(A_\mu) \longrightarrow \L'(A_\mu,\del_\mu B).
$$
It is immediate to note that if we assume the new field transforms
like $\dl B=m\th$, then the requirement of local gauge invariance is
$$
\dl\L' = \biggl[\frac{\del\L'}{\del A_\mu} + m
\frac{\del\L'}{\del(\del_\mu B)}\biggr]\del_\mu\th = 0.
$$
It follows
$$
m\frac{\del\L'}{\del(\del_\mu B)} = - \frac{\del\L'}{\del
A_\mu}.
$$
Consequently only the combination
$$
A_\mu - \del_\mu B/m
$$
occurs in~$\L'(A_\mu,\del_\mu B)$. Thus we rewrite:
$$
\L'(A_\mu,\del_\mu B) \longrightarrow \L_{0,{\rm phmass}}(A_\mu -
\del_\mu B/m).
$$
The bosonic part of the Lagrangian is \textit{in fine}
$$
\L_{\rm b} = \L_{\rm kin} + \frac{m^2}{2}\biggl(A_\mu -
\frac{\del_\mu B}{m}\biggr)^2;
$$
note that, with $V_\mu=(A_\mu - \del_\mu B/m)$, one has $\L_{\rm
kin}(A_\mu)=\L_{\rm kin}(V_\mu)$.  The total Lagrangian~$\L=\L_{\rm
f}+\L_{\rm b}$ has what we want.  With the multiplet of fields
$$
\vf = \begin{pmatrix} \owl\psi \\ \psi \\ A^\mu \\ B \end{pmatrix}
\sepword{transforming as} \dl\vf = \begin{pmatrix} i\owl\psi \th(x) \\
-i\psi \th(x) \\ \del^\mu\th(x) \\ m\th(x) \end{pmatrix},
$$
we plainly obtain local gauge invariance of~$\L_{\rm f},\L_{\rm b}$
and~$\L$.  Note the Euler--Lagrange equation
$$
\del_\mu\frac{\del\L}{\del_\mu B} = \frac{\del\L}{\del B}
\sepword{yielding} \square B = m\,\del A.
$$
Note as well that one can fix the gauge so~$B$ vanishes; this does not
mean the gauge symmetry is trivial.

Maybe Utiyama missed this because~\cite{Ida} he only takes into
account, for the original variables, infinitesimal gauge
transformations typical of `matter' fields, of the
form~\eqref{eq:initial-condition}; he did not consider the
possibility~\eqref{eq:talon-de-Aquiles}, that
is~\eqref{eq:bottom-line}, for the vector fields acting as sources of
gauge~fields.

We finish this subsection by noting that $\L_{\rm b}$ may be written 
as well
$$
\L_{\rm b} = (\del_\mu - igA_\mu)\Phi\,(\del_\mu + igA_\mu)\Phi^*,
\sepword{with} \Phi = \frac{m}{\sqrt2 g}\exp(igB/m);
$$
that is an abelian Higgs model without self-interaction.  The
verification is straightforward.

\subsection{The St\"uckelberg formalism for non-abelian Yang--Mills
fields}

The sophisticated method for this was established by Kunimasa and
Goto~\cite{NewNipponStrike}; we follow in the main~\cite{Banzai}.  For
apparent simplicity, consider an isovector field~$A^a_\mu$ interacting
with an isospinor spinor field~$\psi$, like in subsection~1.5.  Let us
choose the notation
$$
\AA_\mu = \thalf\sigma^aA^a_\mu; \qquad \FF_{\mu\nu} = \del_\mu\AA_\nu
- \del_\nu\AA_\mu + ig(\AA_\mu\AA_\nu - \AA_\nu\AA_\mu).
$$
Indeed $\tiquarter\sigma^a\sigma^b=-\thalf\eps^{abc}\sigma^c$, in
consonance with~\eqref{eq:estas-avisado}.  The Lagrangian density is
written
$$
-\thalf\tr(\FF_{\mu\nu}\FF^{\mu\nu}) +
\tihalf\owl\psi\overleftrightarrow{\del_\mu}\ga^\mu\psi - \owl\psi
M\psi - g\owl\psi \ga^\mu A_\mu\psi.
$$
This is invariant under
$$
\psi \to \W^{-1}\psi; \qquad \AA_\mu \to \W^{-1}\AA_\mu\W - 
\frac{i}{g}\W^{-1}\del\mu\W,
$$
for $\W\in SU(2)$; which is nothing but~\eqref{eq:bottom-line}, with
$$
\W = \exp(T^a\th^a(x)).
$$
To make the mass term
$$
m^2\tr(\AA_\mu\AA^\mu) = \thalf m^2A^a_\mu A^{a\mu}
$$
gauge invariant, it is enough to introduce a $2\x2$ matrix $\om_\mu$
of auxiliary vector fields, so that
$$
m^2\tr(\AA_\mu - \om_\mu/g)
$$
is invariant under gauge transformations, if
\begin{equation}
\om_\mu \to \W^{-1}\om_\mu\W - i\W^{-1}\del_\mu\W.
\label{eq:say-something}
\end{equation}
Indeed, let $C\in SU(2)$ transform as $C\to C\W$.  Then
$$
\om_\mu := -iC^{-1}\del_\mu C
$$
satisfies~\eqref{eq:say-something}:
$$
-i\W^{-1}C^{-1}\del_\mu C\W = \W^{-1}\om_\mu\W -
i\W^{-1}\del_\mu\W.
$$
With $C=\exp(B^aT^a/m)$, we can think of the~$B^a$ as the auxiliary
fields.

\smallskip

We may add, however, that the introduction of scalar St\"uckelberg
partners for the $A^a_\mu$ by the substitution $\AA_\mu\to
\AA_\mu-\del_\mu B$, with $B=B^aT^a$, seems to work as well.  In gauge
theory, the ``elegant'' non-infinitesimal notation is a bit dangerous,
in that it tends to obscure the fact that the transformation of the
gauge fields~\eqref{eq:bottom-line} is \textit{independent} of the
considered representation of the gauge group.

\subsection{Gauge-fixing and the St\"uckelberg Lagrangian}

We begin to face quantization now.  For that, we need to fix a gauge.
Otherwise, we cannot even derive a propagator from the Lagrangian.
Let us briefly recall the standard argument:
$$
\L_{\rm QED} = -\tquarter F_{\mu\nu}F^{\mu\nu} = \thalf
A^\mu\D_{\mu\nu}A^\nu,
$$
with
$$
\D_{\mu\nu}(x) =
-g_{\mu\nu}\overleftarrow{\del_\sigma}\overrightarrow{\del^\sigma} +
\overleftarrow{\del_\nu}\overrightarrow{\del_\mu} \sepword{or}
\D_{\mu\nu}(k) = -g_{\mu\nu}k^2 + k_\mu k_\nu,
$$
in momentum space.  The matrix $\D_{\mu\nu}$ has null determinant and
thus is not invertible; so one cannot define a Feynman propagator.
This is precisely due to gauge invariance.  The problem for QED
was cured by Fermi long ago~\cite{bellacosa} by introduction of the
piece $\frac{-1}{2\a}(\del^\nu A_\nu)^2$.  Here we proceed similarly,
and the gauge-fixing term we take is of the 't~Hooft type:
\begin{equation}
\L_{\rm gf} = \frac{-1}{2\a}(\del^\nu A_\nu + \a mB)^2.
\label{eq:middle-affluent}
\end{equation}
We denote
$$
\L_S = \L + \L_{\rm gf},
$$
the St\"uckelberg Lagrangian.  The gauge-fixing amounts to that now
the gauge variation~$\th$ must satisfy the Klein--Gordon equation with
mass~$m\sqrt\a$:
$$
(\square + \a m^2)\th = 0;
$$
just like in old trick by Fermi in electrodynamics, where the new
Lagrangian is still gauge-invariant provided we assume $\square\th=0$
for the gauge variations.  Now instead the Euler--Lagrange equation
\begin{equation*}
\del_\mu\frac{\del\L_S}{\del(\del_\mu B)} = \frac{\del\L_S}{\del B}
\sepword{yields} (\square + \a m^2)B = 0.
\end{equation*}
Hence the gauge-fixing implies $B$ itself now is a free field with
mass~$m\sqrt\a$.  Another good reason for the gauge-fixing is to keep
$A_\nu$ as an honest-to-God spin~1 field in the interaction.  Recall
that in a quantum vector field spin~0 and~1 are possible.  The
scalar~$B$ `extracts' the spin~0 part, so the remaining part is
transverse.  In fact $\del_\mu(A^\mu-\del^\mu B/m)=\del A+\a mB$ if
the equation of motion is taken into account; and this gauge-fixing
term is destined to vanish in an appropriate sense on the physical
state space.

A word is needed on the Noether theorem now. There is now an extra
term in~$\del_{\mu}\frac{\del\L} {\del(\del_{\mu}A_{\nu})}$, of the
form $-\frac{g^{\mu\nu}}{\a}(\del A + \a mB)$. This gives rise to the
Euler--Lagrange equation:
\begin{equation}
\square A_\mu + \Bigl(\frac{1}{\a}-1\Bigr)\del_\mu(\del A) + m^2A_\mu
= g\owl\psi\ga_\mu\psi,
\label{eq:harbinger-of-trouble}
\end{equation}
where we have restablished temporarily the coupling constant. As a
consequence of~\eqref{eq:harbinger-of-trouble} we have
\begin{equation*}
\square\,\del A + \a m^2\del A = 0.
\end{equation*}
The simplest option now is to take $\a=1$ (so the masses of $A_\nu$
and $B$ coincide), as then the $A_\nu$ obey the Klein--Gordon equation
at zeroth order in~$g$.  This could be termed the `Feynman gauge'.
But in some contexts it is important to keep the freedom of different
mass values for the vector and the scalar bosons.  (We have for the
fermion the Dirac equation
$$
i\ga^\mu\del_\mu\psi = (g\ga^\mu A_\mu + M)\psi,
$$
and its conjugate. Nothing new here.)

A comment on renormalizability is in order at this point.  The choice
$\a\downarrow0$ is the Landau gauge, in which renormalizability is
almost explicit.  On the other hand, it is clear that $B=0$ (the
original Proca model), where the theory is non-renormalizable by power
counting, can be recovered as a sort of `unitary gauge'.  If we can
prove gauge covariance of the theory, all these versions will be
physically equivalent.  An extra advantage of the St\"uckelberg field
in renormalization is that, because it cures the limit $m\downarrow0$,
it allows the use of masses as infrared regulators.

\smallskip

To finish, we call the attention again upon the similitudes of the
model with the abelian Higgs model.  Upon renormalization, a ``Higgs
potential-like'' term pops up in the Lagrangian.  However, the vacuum
expected value of the St\"uckelberg field is still zero.  For
non-abelian theories, the situation remains murky even now.

\subsection{The ghosts we summoned up}

For completeness, we insert next a conventional discussion of BRS
invariance for the Lagrangian obtained in the previous subsection.
(This is not intended to be discussed during the lessons, and both the
cognoscenti and the non-cognoscenti may skip it in first reading.)

Nowadays BRS invariance of the (final) Lagrangian is an integral part
of the quantization process. Among other things, it helps to establish
gauge \textit{covariance}, that is, independence of the chosen gauge
for physical quantities; in turn this helps with renormalizability
proofs. We approach the quantum context by introducing two fermionic
ghosts $\om,\omf$ plus an auxiliar (Nakanishi--Lautrup) field~$h$ that
we add to the collection~$\vf$. From the infinitesimal gauge
transformations we read off the BRS transformation:
\begin{equation*}
s\vf = s\begin{pmatrix} \owl\psi \\ \psi \\ A^\mu \\ B \\ \om \\ \omf
\\ h \end{pmatrix} = \begin{pmatrix} i\om\owl\psi \\ -i\om\psi \\
\del^\mu\om \\ m\om \\ 0 \\ h \\ 0 \end{pmatrix}.
\end{equation*}
It is clear that $s$ increases the ghost number by~one. Extend $s$ as
an antiderivation; from the fact that $\om,\omf$ are anticommuting we
obtain (even off-shell) nilpotency of order two for the BRS
transformation: $s^2=0$ (we will always understand `nilpotent of order
two' for `nilpotent' in this work). Now, in the BRS approach, one
takes the action to be a local action functional of matter, gauge,
ghost and $h$-fields with ghost number zero and invariant under~$s$.
This is provided by the new form
$$
\L_{\rm gf} = s[\F(\owl\psi,\psi,A_\mu,B)\omf + \thalf\a h\omf],
$$
for the gauge-fixing term of the Lagrangian.  Here $\F$ is the
gauge-fixing functional, like $(\del^\mu A_\mu + \a mB)$ above.
Invariance comes from $s\L_{\rm gf}=0$ on account of nilpotency, of
course.  We can rewrite
$$
\L_{\rm gf} = -\omf s\F + h\F + \thalf\a h^2 = -\omf s\F +
\thalf\Bigl(\frac{\F}{\sqrt\a} + h\sqrt\a\Bigr)^2 - \frac{\F^2}{2\a}.
$$
One can eliminate $h$ using its equation of motion
$$
0 = \frac{\partial\L_{\rm gf}}{\partial h} = \F + h\a, \sepword{so
that} \L_{\rm gf} = - \omf s\F -\frac{\F^2}{2\a}.
$$
and also $s\omf = -\F/\a$: the BRS transformation maps then the
anti-ghosts or dual ghosts into the gauge-fixing terms (the price to
pay is that $s$ would be nilpotent off-shell only when acting on
functionals independent of~$\omf$). In our
case~\eqref{eq:middle-affluent}:
$$
s\F = s(\del^\mu A_\mu + \a mB) = (\square + \a m^2)\om.
$$
Thus the contribution of the fermionic ghosts in this abelian model
to~$\L_{\rm gf}$ is
$$
-\omf s\F = -\omf(\square + \a m^2)\om;
$$
also $\del_\mu\omf\del^\mu\om-\omf\a m^2\om$ would do; the ghosts turn
out to be free fields with the same mass as St\"uckelberg's~$B$-field.
Notice that the ghost term decouples in the final effective
Lagrangian.  (According to~\cite{mavericks}, adding to the action a
term invariant under the BRS transformation amounts to a redefinition
of the fields coupled to the source in the generating functional; this
has no influence on the~$\Sf$-matrix.)

We have followed~\cite{VacuumCleanerBis} and mainly~\cite{UpsideDown}
in this subsection.

\smallskip

At the end of the day, the Lagrangian for massive electrodynamics
is of the form
\begin{align*}
&\L_{\rm f} + \L_{\rm kin} + \L_{\rm b} + \L_{\rm gf} =
\tihalf[\owl\psi\ga^\mu\del_\mu\psi - \del_\mu\owl\psi\ga^\mu\psi] -
\owl\psi\ga^\mu A_\mu\psi - \owl\psi M\psi
\\
& -\tquarter(FF) + \frac{m^2}{2}(A - \del B/m)^2 - \frac{1}{2\a}(\del
A + \a mB)^2 -\omf(\square + \a m^2)\om
\\
&= \L_{\rm f} + \L_{\rm kin} + \frac{m^2A^2}{2} - \frac{1}{2\a}(\del
A)^2 + \thalf(\del B)^2 - \frac{\a m^2}{2}B^2 - m\del_\mu(BA^\mu)
\\
&= -\omf(\square + \a m^2)\om.
\end{align*}

Highlights:

\begin{itemize}
\item{}
The gauge-fixing has been chosen independently of the matter field. 

\item{}
The gauge sector contains first a massive vector field, with three 
physical components of mass~$m$ (one longitudinal and two transverse)
and an unphysical spin-zero piece of mass~$\sqrt\a m$.

\item{}
The cross term between $A_\mu$ and~$B$ disappeared.

\item{}
The gauge sector also contains a (commuting) St\"uckelberg $B$-field
with mass~$\sqrt\a m$ and a pair of (anticommuting) ghost-antighost
scalars, with mass~$\sqrt\a m$ as well.

\item{}
For computing $\Sf$-matrix elements, the ghosts can be integrated out,
since they are decoupled and do not appear in asymptotic states.  But
we cannot integrate out the $B$-field, because, as discussed in
Section~3, it plays a role in the definition of the physical states 
---and moreover it undergoes a non-trivial renormalization.

\item{}
The only interacting piece is the $\owl\psi A\psi$ term in the
fermionic part of the Lagrangian.

\item{}
The model is \textit{renormalizable}.
\end{itemize}

\section{Quantization of massive spin-1 fields}

\subsection{On the need for BRS invariance}

It is impossible for us, within the narrow limits of this short
course, to follow in any meaningful detail the tortuous chronological
path to the discovery of BRS invariance in relation with gauge
invariance. The story in outline is well-known. By fixing the gauge,
Feynman was able to generate Feynman diagrams~\cite{Ulises} for
non-abelian gauge theories; but unitarity of the $\Sf$-matrix was lost
unless additional ``probability-eating'' quantum fields were
introduced. The auxiliary ghost fields appeared clearly in the work by
Fadeev and Popov, that uses the functional integral. In the seventies
it was discovered that the resulting effective Lagrangian still
supports a global invariance of a new kind, the nilpotent BRS
transformation, that allows to recover unitarity, ensures gauge
independence of the quantum observables and powerfully contributes to
the proofs of renormalizability.

We attacked quantization in subsection~2.1 through the canonical
method.  So we motivate the introduction of the ghosts and BRS
symmetry/operator in our previous considerations.  Now that hopefully
we have broken the mental association between ``gauge principle'' and
``masslessness'', one can proceed to a simple and general version of
gauge theory with BRS invariance.  The quantization of massive vector
fields is interesting in that it is conceptually simpler, although
analytically more complicated, than that of massless ones.  (It is
true that in theories with massive gauge bosons, the masses are
generated by the `Higgs mechanism'; but this is just a poetic
description that cannot be verified or fasified at present.)  In the
context, concretely we need the ghosts as ``renormalization
catalysers''.  In fact, it has been shown in~\cite{CabezondelaSal}
that for interacting massive vector field models the renormalizability
condition fixes the theory completely, including the cohomological
extension of the Wigner representation theory by the ghosts, and the
St\"uckelberg field in the abelian case ---even if you had never heard
of it in a semi-classical study of Lagrangians, like the one performed
in Section~2.  As well as a Higgs-like field for flavourdynamics; we
shall touch upon this in the last section.

The crucial problem, illustrated by our discussion in subsection~2.1,
is to eliminate the unphysical degrees of freedom in the quantization
of free vector fields in a subtler way than Proca's, particularly
without giving up commutators of the form
\begin{equation}
[A_\mu(x), A_\nu(y)] = ig_{\mu\nu}D(x-y), \quad A_\mu^{+} = A_\mu.
\label{eq:hold-your-breath}
\end{equation}
Also we ask for the KG equations $(\square + m^2)A^\mu=0$ to hold (in
the Feynman gauge).  It is impossible to
realize~\eqref{eq:hold-your-breath} on Hilbert space.  Let us sketch
the solution in this subsection.  It goes through the introduction of
a distinguished symmetry~$\eta$ (that is, an operator both selfadjoint
and unitary), called the Krein operator, on the Hilbert--Fock
space~$H$.  Whenever such a Krein operator is considered, the
$\eta$-adjoint~$O^+$ of an operator~$O$ is defined:
$$
O^+ = \eta O^\7\eta.
$$
Let~$(.,.)$ denote the positive definite scalar product in~$H$. Then
$$
\<.,.> := (.,\eta.)
$$
gives an `indefinite scalar product', and the definition of~$O^+$ is
just that of the adjoint with respect to~$\<.,.>$. The algebraic
properties are like in usual adjugation~$\7$, but $O^+O$ is not
positive in general.

The pair $(H,\eta)$, where $H$ is the original Hilbert--Fock space,
including ghosts, is called a Krein space. The undesired contributions
from the $A$-space will be cancelled by the `unphysical' statistics of
the ghosts. The BRS operator is an (unbounded) nilpotent
$\eta$-selfadjoint operator~$Q$ on~$H$. That is, $Q^2=0,Q=Q^+$. By
means of~$Q$ one shows that $H$ (or a suitable dense domain of it)
splits in the direct sum of three pairwise orthogonal subspaces (quite
analogous to the Hodge--de Rham decomposition in differential geometry
of manifolds):
$$
H = \ran Q \op \ran Q^\7 \op (\ker Q\cap\ker Q^\7).
$$
In addition we assume
$$
\eta\Big|_{\ker Q\cap\ker Q^\7} = 1.
$$
That is, $\<.,.>$ is positive definite on
$$
H_{\rm phys} := \ker Q\cap\ker Q^\7,
$$
which is called the physical subspace. An alternative definition 
for~$H_{\rm phys}$ is the cohomological one:
$$
H_{\rm phys} = \ker Q/\ran Q.
$$
Nilpotency of~$Q$ is the reason to introduce the anticommuting pair of
ghost fields. In interaction, the $\Sf$-matrix must be physically
consistent:
$$
[Q, \Sf]_+ = 0, \sepword{or at least} [Q, \Sf]_+\Big|_{\ker Q} = 0.
$$
In the following subsections we flesh out the details of all this.

\subsection{Ghosts as free quantum fields}

A first step in a rigorous construction of ghosts is their
understanding as quantum fields, together with the issue of the
`failure' of the spin-statistics theorem for them.  We look for two
operator-valued distributions $u,\ut$, acting on a Hilbert--Fock
space~$H_{\rm gh}$ and satisfying Klein--Gordon (KG) equations:
\begin{equation}
(\square + m^2)u = (\square + m^2)\ut = 0,
\label{eq:jaque-teresiano}
\end{equation}
and the following commutation relations, in the sense of tempered
distributions
$$
[u_a(x), \ut_b(y)]_+ = -i\dl_{ab}D(x-y); \qquad [u_a(x), u_b(y)]_+ =
[\ut_a(x), \ut_b(y)]_+ = 0.
$$
Here $D=D^++D^-$ is the Jordan--Pauli function; we refer to the
supplement at the end of these notes for notation regarding the
propagators.  The fields `live' in the adjoint representation of a
gauge group~$G$ (as the gauge fields themselves); the colour
indices~$a,b$ most often can be omitted.  The components of~$H_{\rm
gh}$ of degree~$n$ are \textit{skewsymmetric} square-summable
functions (with the Lorentz-invariant measure $d\mu_m(p)$) of~$n$
momenta on the mass hyperboloid~$\H_m$, with their colour indices and
ghost indices, where the first, say~$a$, can run from 1 to~$\dim G$,
and we let the second, say~$i$, take the values~$\pm1$.  (The reader
is warned of that the notation for the ghost fields in this section,
and a few other notational conventions, are different from we found
convenient in the sections dealing with the semi-classical aspects.)

We proceed to the construction. Consider the dense domain $\D\subset
H_{\rm gh}$ of vectors with finitely many nonvanishing components
which are Schwartz functions of their arguments. Then there exist the
annihilation (unbounded) operator functions $c_{a,i}(p)$ of~$\D$ into
itself, given by
$$
[c_{a,i}(p)\Phi]^{(n)}_{a_1,\ldots,a_n;i_1,\ldots,i_n}(p_1,\ldots,p_n)
= \sqrt{n+1}\,
\Phi^{(n+1)}_{a,a_1,\ldots,a_n;i,i_1,\ldots,i_n}(p,p_1,\ldots,p_n).
$$
Integrating this with a Schwartz function on the mass hyperboloid
gives a bounded operator. The adjoint of~$c_{a,i}(p)$ is defined as a
sesquilinear form on~$\D\ox\D$, and we have the usual ``commutation
relations'' among them:
$$
[c_{a,i}(p), c^\7_{b,j}(p')]_+ = \dl_{ab}\dl_{ij}\dl(p-p');
$$
otherwise zero.  Notice that $\dl(p-p')$ is shorthand for the Lorentz
invariant Dirac distribution $2E\dl(\vec p-\vec p')$ corresponding
to~$d\mu_m(p)$.

We are set now to define the distributional ghost field operators in
coordinate space out of the $c_{a,i},c^\7_{b,j}$. The construction is
diagonal in the $G$-index, so it will be omitted. The general Ansatz
is
$$
u_i(x) = \int d\mu_m(p)\,\big[A_{ij}c_j(p)e^{-ipx} +
B_{ij}c^\7_j(p)e^{+ipx}\big].
$$
Here
$$
A = \begin{pmatrix} A_{11} & A_{1-1} \\ A_{-11} &
A_{-1-1}\end{pmatrix}; \qquad B = \begin{pmatrix} B_{11} & B_{1-1} \\
B_{-11} & B_{-1-1}\end{pmatrix}.
$$
Since $p$ is on the mass hyperboloid the KG
equations~\eqref{eq:jaque-teresiano} hold. The anticommutators are:
$$
[u_i(x), u_j(y)]_+ = -i\big[A_{ik}B_{jk}D^+(x-y) -
B_{ik}A_{jk}D^-(x-y)\big].
$$
The only combinations with causal support are multiples of $D^++D^-$.
As we want to keep causality, it must be $AB^t+BA^t=0$, so we obtain
$$
[u_i(x), u_j(y)]_+ = -iC_{ij}D(x-y),
$$
with $C:=AB^t$ skewsymmetrical. There are of course many possible
choices of~$A,B$ with this constraint. We pick:
$$
C = \begin{pmatrix} & 1 \\ -1 & \end{pmatrix}.
$$
This finally gives:
\begin{align*}
u(x) = u_1(x) &= \int d\mu_m(p)\,\big(c_1(p)e^{-ipx} +
c^\7_{-1}(p)e^{ipx}\big);
\\
\ut(x) = u_{-1}(x) &= \int d\mu_m(p)\,\big(c_{-1}(p)e^{-ipx} -
c^\7_1(p)e^{ipx}\big).
\end{align*}
We remark $[\ut(x), u(y)]_+=iD(x-y)=-iD(y-x)=[u(y), \ut(x)]_+$.

\smallskip

The representation of the Poincar\'e group is the same as for $2\dim
G$ independent scalar fields; we do not bother to write it. As we have
chosen $A,B$ invertible, the creation and annihilation operators can
be expressed in terms of the ghost fields and their adjoints. Then the
vacuum is cyclic with respect to these.

Defining the adjoint fields, one sees that the anticommutators of the
ghost fields with their adjoints are not causal. This, according
to~\cite{Kraken,ETHneverdies} allows to escape the spin-statistics
theorem. Indeed, a version of the last says that no nonvanishing
scalar fields can exist satisfying
$$
[u_a(x), u_b(y)]_+ = 0, \qquad [u_a(x), u_b^\7(y)]_+ = 0,
$$
for spacelike separations.  Because the second anticommutator is not
causal, the last condition is not violated.  (There are other
explanations in the literature for the same conumdrum, though.)

\subsection{Mathematical structure of BRS theories}

There are several questions relative at the scheme proposed in~3.1,
that we address systematically now.

\begin{enumerate}
\item{} What is the algebraic framework?

\item{} In which mathematical sense BRS invariance is a symmetry?

\item{} When is there a BRS charge associated to a BRS symmetry?

\item{} What are the continuity properties of the generator~$Q$?

\item{} How the `Hodge--de Rham' decomposition of the Hilbert space
takes place?

\item{} How are the physical states characterized?
\end{enumerate}

The first paper to tackle these questions was the famous on the quark
confinement problem by Kugo and Ojima~\cite{AlwaysNippon}, although
their answers were not quite correct.  A very good treatment, that we
follow for the most part, was given by Horuzhy and
Voronin~\cite{SteklovRomance}.

\begin{enumerate}
\item{} Consider a `general BRS theory' on a Krein space $(H,\eta)$.
On a suitable common invariant dense domain $\D\subset H$ there is
defined a system of physical quantum fields and ghost fields (the
physical fields could be matter fields, Yang--Mills fields or, say,
the coordinates of a first-quantized string), forming a polynomial
algebra~$\A$; the operator $\id\in\A$ on~$H$ we denote by~1. A Krein
operator has the eigenvalues~$\pm1$, so $\eta=P^\eta_+-P^\eta_-$ with
an obvious notation. We assume moreover $\dim P^\eta_{\pm}H=\infty$.
By~$O^\circ$ we shall mean the restriction of~$O^+$ to~$\D$. We say
$O$ is $\eta$-selfadjoint when $O=O^\circ$; $\eta$-unitary when
$O^{-1}=O^\circ$. The field algebra has a cyclic vector or
`vacuum'~$\vac$, that is, $\A\vac$ is dense in~$\D$.

\item{} Mathematically speaking, a BRS (infinitesimal) transformation
is a skew-adjoint, nilpotent superderivation $s$ acting on the field
algebra of~$H$. Let $\eps_O:=(-)^{\N_{\rm gh}(O)}$, whith $\N_{\rm
gh}(O)$ the number of ghost fields in the monomial~$O$. Typically $s$
changes the ghost number by one. Then $s$ is a linear map of~$\A$
into~$\A$ such that
\begin{align*}
s(OB) &= s(O)B + \eps_ OOs(B), \; s^2=0, \; \eps_{s(O)} = -\eps_O
\\
&\sepword{and} s(O)^\circ = -\eps_Os(O^\circ).
\end{align*}
The key point for BRS invariance is obviously the nilpotency equation
$s^2=0$.

\item{} An important question is whether the BRS transformation~$s$
possesses a generator or BRS charge~$Q$, that is, takes the form
\begin{equation}
s(O) = [Q, O]_{\pm} \sepword{where} [Q, O]_{\pm} := QO - \eps_O OQ.
\label{eq:horror-y-pavor}
\end{equation}
Indeed, we may try to equivalently write~\eqref{eq:horror-y-pavor} as
\begin{equation*}
QO\vac = s(O)\vac.
\end{equation*}
This equation will serve as definition of~$Q$, at least on a dense
subset of~$\D$, provided
$$
O\vac = 0 \sepword{implies} s(O)\vac = 0.
$$
Note that $Q\vac=0$ because $s(1)=0$. Thus~\eqref{eq:horror-y-pavor}
is consistent. Nilpotency of~$Q$ follows:
$$
Q^2O\vac = Qs(Q)\vac = 0.
$$
One expects $Q$ as defined above to be $\eta$-selfadjoint. But this is
not completely automatic. We have
\begin{align}
\<QO\vac, B\vac> &= \<s(O)\vac, B\vac> = \<B^\circ s(O)\vac, \vac>
\nonumber \\
&= \eps_{B^\circ}\<\big(s(B^\circ O) - s(B^\circ)O\big)\vac, \vac>
\nonumber \\
&= \eps_O\vev{s(O^\circ B)}
\nonumber \\
&+ \<O\vac, s(B)\vac>.
\label{eq:truculent}
\end{align}
This will be equal to $\<O\vac, QB\vac>$ if in general we have
$$
\vev{s(O)} = 0 \sepword{for all} O\in\A.
$$
In this case, we have $\eta$-symmetry. For passing to 
$\eta$-selfadjointness, consult~\cite{DonAlberto}.

Reciprocally, if $Q$ is $\eta$-selfadjoint with~$Q\vac=0$, nilpotent,
and generates~$s$ by~\eqref{eq:horror-y-pavor}, then, rather
trivially:
$$
\vev{s(O)} = \vev{[Q, O]_{\pm})} = \<Q\vac,O\vac> = 0.
$$
Moreover, for $s$ so defined
\begin{align*}
s(O)^\circ &= (QO - \eps_O OQ)^\circ = O^\circ Q - \eps_O QO^\circ
\\
&= - \eps_O(QO^\circ - O^\circ Q) = -\eps_Os(O^\circ).
\end{align*}
We finally verify nilpotency of~$s$:
\begin{equation*}
s^2(O) := [Q, [Q, O]_{\pm}]_{\pm} = Q(QO - \eps_O OQ) + \eps_O(QO - 
\eps_O OQ)Q = 0.
\end{equation*}

\item{} In physics $Q$ is often treated as a bounded operator. But
there are large classes of nilpotent, $\eta$-selfadjoint unbounded
operators. Let for instance $H=H_1\op H_2$ and
$$
\eta=\begin{pmatrix} 1 & 0 \\ 0 & -1 \end{pmatrix} \sepword{with}
Q=\begin{pmatrix} 0 & A \\ 0 & 0 \end{pmatrix},
$$
with $A$ unbounded and skewadjoint.  Then $Q$ is nilpotent,
$\eta$-selfadjoint and unbounded.  For another example, take $H=H_1\ox
H_2$, where $H_1$ is an infinite-dimensional Hilbert space, $H_2$ is a
Krein space, $Q=O\ox B$, with $O=O^\7$ unbounded and~$B$ nilpotent and
$\eta$-selfadjoint.  Typically BRS operators are sums of such
operators.

Given an arbitrary nilpotent operator~$Q$, such that $\dom Q^2$ is
dense, the following holds: either $Q$ is bounded, with~$0$ as unique
point in its spectrum, or $Q$ is unbounded and its spectrum is all of
the complex plane.

\begin{proof}
Assume $\spec Q\ne\C$. Let $\la$ belong to the resolvent of~$Q$. Then
$Q$ is closed, as $Q-\la$ is. (We recall that a Hilbert space operator
is by definition closed when its graph is closed. Also by definition,
$Q-\la$ is a one-to-one map from $\dom Q$ onto~$H$ with bounded
inverse, so it is closed.) Now $(Q-\la)^{-1}H\subset\dom Q$. Therefore
$$
(Q - \la)\big(Q + \la Q(Q - \la)^{-1}\big)
$$
makes sense and is equal to~$Q^2$. Now $Q$ is closed and $\la
Q(Q-\la)^{-1}$ is bounded, therefore $Q + \la Q(Q - \la)^{-1}$ is
closed; then $Q^2$ is closed. Therefore its domain is all of~$H$, so
$Q$ is bounded (by the closed graph theorem). Then it is well known
that~$\spec Q=\set{0}$.
\end{proof}

\item{} Consider the subspaces $\ker Q,\eta\ker Q,\ran Q,\eta\ran Q$.
Due to $Q^2=0$, we can assume $\ran Q\subset\dom Q$; otherwise we
extend $Q$ to the whole $\ran Q$ by zero. Because of
$\eta$-selfadjointness, $\ker Q$ is closed; also, $\eta\ran Q=
\eta\ran\eta Q^\7\eta=\ran Q^\7$ and $\eta\ker Q=\ker\eta Q\eta=\ker
Q^\7$. In view of nilpotency, it is immediate that
$$
\ran Q \perp \ran Q^\7,
$$
where $\perp$ indicates perpendicularity in the Hilbert space sense.
We have
$$
(\ran Q \op \ran Q^\7)^\perp = \ker Q^\7 \cap \ker Q.
$$
Indeed, the domain of~$Q^\7$ is dense in~$H$ and thus $(x,Q^\7y)=0$
for all $y\in\dom Q^\7$ implies $Qx=0$. Similarly for $(\ran
Q)^\perp=\ker Q^\7$. Denoting by $[\perp]$ perpendicularity in the
Krein space sense, it also clear that
$$
(\ker Q^\7\cap\ker Q)^\perp = (\ker Q^\7\cap\ker Q)^{[\perp]} 
$$
In summary
\begin{align}
H &= \overline{\ran Q^\7} \op \ker Q = \overline{\ran Q} \op \ker Q^\7
= \overline{\ran Q} \op \overline{\ran Q^\7} \op(\ker Q^\7\cap\ker Q)
\nonumber \\
&= \overline{\ran Q} \op \overline{\ran Q^\7} [+] (\ker Q^\7\cap\ker
Q);
\label{eq:yours-truly}
\end{align}
where the last symbol means the $\eta$-orthogonal sum. This is the
`Hodge--de Rham' decomposition of~$H$.

\item{} 
Assume moreover
$$
\eta\Big|_{\ker Q\cap\ker Q^\7} = 1.
$$
Then we baptize
$$
H_{\rm phys} := \ker Q\cap\ker Q^\7,
$$
the physical subspace, on which $\<.,.>$ is positive. Alternative 
characterizations are
$$
H_{\rm phys} = \ker Q/\overline{\ran Q},
$$
in view of~\eqref{eq:yours-truly}, and
$$
H_{\rm phys} = \ker[Q, Q^\7]_+.
$$
Indeed $[Q, Q^\7]_+\,x=0$ iff $Qx=Q^\7x=0$.
\end{enumerate}

\subsection{BRS theory for massive spin one fields}

We finally turn to our physical case. When dealing with the massive
vector field, instead of eliminating ab initio the longitudinal
component as in~\eqref{eq:in-articulo-mortis}, we keep the $a(k,0)$
and their adjoints, and proceed as follows. We recognize Krein spaces
as appropriate tools to study (quantum) gauge theories. In our present
case $\eta:=(-)^{\N_l}$, where $\N_l$ is the particle number operator
for the longitudinal modes. Now
\begin{equation*}
A^\mu(x) = (2\pi)^{-3/2}\sum_{\sigma=0}^3 \int d\mu_m(k)\,
\big(\eps^\mu(k,\sigma)e^{-ikx}a(k,\sigma) +
\eps^\mu(k,\sigma)e^{ikx}a^+(k,\sigma)\big).
\end{equation*}
Clearly
$$
a^+(k,0) = -\eta^2a^\7(k,0) = -a^\7(k,0);
$$
however, by definition $A^\mu(x)$ is $\eta$-selfconjugate.

\smallskip

We hasten to indicate the main difference with the massless case. Note
that a unitary representation of the Poincar\'e group on the original
space is given by
$$
U(a,\La)A^\mu(x)U^{-1}(a,\La) = \La^\mu_\nu A^\nu(\La x + a) =
U^{-1+}(a,\La)A^\mu(x)U^+(a,\La).
$$
This implies
$$
[U^+(a,\La)U(a,\La), A^\mu(x)] = 0;
$$
therefore $U$ is $\eta$-unitary. As $\N_l$, thus~$\eta$, commutes
with~$U$ ---basically because the longitudinal polarization transforms
into itself under a Lorentz transformation,
$$
\La^\nu_\mu\eps^\mu(k,0) = \frac{(\La k)^\nu}{m} = \eps^\nu(\La k,0),
$$
the representation~$U$ is also unitary. This cannot be obtained in the
massless case.

\smallskip

The commutation relations for $A$-field are of the form
$$
[A^\mu(x), A^\nu(y)] = ig^{\mu\nu}D(x-y),
$$
as we wished for. We now employ a nilpotent gauge charge~$Q$ to
characterize the physical state subspace and eliminate the unphysical
longitudinal mode. For photons, the definition of~$Q$ is known to be
\begin{equation}
Q = \int_{x^0={\rm const}}d^3x\,(\del\.A)\overleftrightarrow{\del_0}u.
\label{eq:amargo-trago}
\end{equation}
Let us accept this is a conserved quantity, associated to the current
$$
j_\mu = (\del\.A)\overleftrightarrow{\del_\mu}u.
$$
Obviously $[Q,u]=0$. By use of the algebraic identity
$$
[AB, C]_+ = A[B, C]_+ - [A, C]B,
$$
nilpotency then is checked as follows:
$$
2Q^2 = [Q, Q]_+ = -\int_{x^0={\rm const}}d^3x\,
[(\del\.A),Q]\overleftrightarrow{\del_0}u = i\int_{x^0={\rm const}}d^3x\,
\square u\overleftrightarrow{\del_0}u = 0,
$$
because the ghost is a free massless quantum field, ie, satifies the
wave equation. The form~\eqref{eq:amargo-trago} will not do for the
massive case, as now, with ghost fields of the same mass as~$A^\mu$,
after a relatively long calculation involving the solution of the 
Cauchy problem for~$u$, we would obtain
$$
2Q^2 = i\int_{x^0={\rm const}}d^3x\,\square u
\overleftrightarrow{\del_0}u = -im^2\int_{x^0={\rm const}}d^3x\,u
\overleftrightarrow{\del_0}u \ne 0,
$$
A suitable form of~$Q$ is reached by introducing a (Bose) scalar field
with the same mass, satisfying
$$
(\square + m^2)B = 0, \quad [B(x), B(y)] = -iD(x-y),
$$
and then
\begin{equation}
Q = \int_{x^0={\rm const}}d^3x\,(\del\.A + mB)
\overleftrightarrow{\del_0}u.
\label{eq:menos-amargo-trago}
\end{equation}
We leave to the care of the reader to check this is a conserved
quantity. Now we obtain
$$
2Q^2 = i\int_{x^0={\rm const}}d^3x\,\square u
\overleftrightarrow{\del_0}u + im^2\int_{x^0={\rm const}}d^3x\,u
\overleftrightarrow{\del_0}u = 0.
$$
In this way we have recovered the St\"uckelberg field!

In summary, the gauge variations are:
\begin{align}
sA^\mu(x) &= [Q, A^\mu(x)]_\pm = i\del^\mu u(x);
\nonumber \\
sB(x)     &= [Q, B(x)]_\pm     = imu(x);
\nonumber \\
su(x)     &= [Q, u(x)]_\pm     = 0;
\nonumber \\
s\ut(x)   &= [Q, \ut(x)]_\pm   = -i\big(\del^\mu A_\mu(x) + mB(x)\big);
\label{eq:fearsome-departure}
\end{align}
with respect to the semi-classical analysis in Section~2 there is a
slight change of notations; the present ones are more advantageous
when dealing with quantum fields. As expected, the BRS variation
of the gauge field corresponds to susbtituting the ghost field for 
the infinitesimal parameter of the gauge transformation.

We finish by a little collection of remarks.

\begin{itemize}
\item{} The ghost number of~$Q$ is precisely~1.

\item{} In view of nilpotency of $Q$, finite gauge variations are
easily computed. We have
$$
A'_\mu(x) = e^{-i\la Q}A_\mu(x)e^{i\la Q} = A_\mu(x) - i\la[Q,
A_\mu(x)] - \thalf\la^2[Q, [Q, A_\mu]].
$$
Note that the last term is \textit{not} zero. But certainly there are 
no higher-order terms.

\item{} Only unphysical fields appear in the
formula~\eqref{eq:menos-amargo-trago} for~$Q$.
	
\item{} A stronger BRS theory includes the anti-BRS symmetry $\bar s$,
with the `complete nilpotency' conditions $s^2={\bar s}^2=s\bar s+
{\bar s}s=0$~\cite{PowerfulSpirits}. The main role of~$\bar s$ is to
ensure the closure of the classical algebra, at the level of
Lagrangians. This is more or less unnecessary in Yang--Mils theories,
but useful for instance in supersymmetric theories.

\item{} It would seem that the foregoing analysis applies only to
abelian fields.  The cognoscenti would in general expect in
formula~\eqref{eq:fearsome-departure} extra terms in the first
equality (covariant derivative rather than ordinary one) and in the
the third one (a ghost term involving the structure constants).  That
is:
\begin{align}
sA^a_\mu(x) &= [Q, A^a_\mu(x)] = iD_\mu u^a(x);
\nonumber \\
su^a(x)     &= [Q, u^a(x)]_+   = -\tihalf gf^{abc}u^b(x)u^c(x);
\label{eq:vuelta-al-redil}
\end{align}

However, it ain't necessarily so.  By just adding the colour index,
one can think of~\eqref{eq:fearsome-departure} as a first step, one in
which self-interaction is neglected, for a non-abelian theory.  In the
causal approach to QFT~\cite{ETHneverdies}, one approaches interacting
fields by means of free fields, and then both methods differ.  
\end{itemize}

\subsection{The ghostly Krein operator}

For completeness, we include here a discussion on the ``charge
algebra'' for ghosts.  Let $f_r$ denote an orthonormal basis
of~$L^2(\H_m,d\mu_m(p))$.  Consider the charge operators
$$
Q(A) := \sum_{r,b,i}c^\7_{b,i}(f_r)a_{ij}c_{b,j}(f_r) = \sum_{b,i}\int
d\mu_m(p)\,c_{b,i}^\7(p)a_{ij}c_{b,j}(p),
$$
for $A=(a_{ij})$ a 2$\x$2 matrix. This is defined on a common dense
domain of~$H_{\rm gh}$, bigger than~$\D$, which is mapped by the
charge operators into itself. This map represents $\gl(2,\C)$, as
$$
Q(AB-BA) = Q(A)Q(B) - Q(B)Q(A); \sepword{also} Q(A^\7) = Q^\7(A).
$$
By the way, by $Q^\7(A)$ we mean its restriction to~$\D$. Taking
for~$A$ the unit matrix and the Pauli matrix~$\sigma_3$, we
respectively obtain the ghost number~$\N_{\rm gh}$ and ghost
charge~$Q_{\rm gh}$ operators. The other two Pauli matrices yield
ghost-antighost exchanging operators, respectively called here
$\Ga,\Om$. Their commutators with the local fields $u,\ut$ are:
\begin{align*}
[\N_{\rm gh}, u] &= -\ut^\7, \qquad [\N_{\rm gh}, \ut] = u^\7;
\\
[Q_{\rm gh}, u] &= -u, \qquad [Q_{\rm gh}, \ut] = \ut;
\\
[\Ga, u] &= \ut, \qquad [\Ga, \ut] = u;
\\
[\Om, u] &= -i\ut, \qquad [\Om, \ut] = iu.
\end{align*}
The verification of this is an exercise. The generator of~$\N_{\rm
gh}$, that constitutes the centre of the charge algebra, gives by
commutation with $u,\ut$ not relatively local fields. We write down
the following currents:
\begin{align*}
j_{\N_{\rm gh}}(x) &:=
i\wick:u^\7(x)\overleftrightarrow{\del^\mu}u(x):; \quad j_{\rm gh}(x)
:= i\wick:\ut(x)\overleftrightarrow{\del^\mu}u(x):;
\\
j_u(x) &:= i\wick:u(x)\overleftrightarrow{\del^\mu}u(x):; \quad
j_\ut(x) := i\wick:\ut(x)\overleftrightarrow{\del^\mu}\ut(x):.
\end{align*}
Again, $j_{\N_{\rm gh}}$ is not a relatively local quantum field. They
are related to the corresponding charges in the usual way; one has,
moreover
$$
\Ga = \thalf(Q_u - Q_\ut), \quad \Om = \tihalf(Q_u + Q_\ut).
$$
We can consider as well operators $T\big(e^{iA}\big):=\exp(iQ(A))$.
They give a representation of the general linear group. It is
$T(B^\7)=T(B)^\7$. Also $T(B)Q(A)T^{-1}(B)=Q(BAB^{-1})$.

\smallskip 

The theory with ghosts has to be constructed by using only the fields
$u,\ut$, while their adjoints will not appear at all; in this way the
troubles with locality are avoided. In massless Yang--Mills theories,
say, one considers the interaction
\begin{equation}
T_1(x) = \tihalf f^{abc}\big(\wick:A^a_\mu A^b_\nu F^{c\mu\nu}:(x)
+ \wick:A^a_\mu u^b\del^\mu\ut^c:\big)(x).
\label{eq:summacumlaude}
\end{equation}
This is invariant under gauge transformations generated by the
differential operator~\eqref{eq:amargo-trago}. The $u^\7,\ut^\7$ do
not appear here. But then it is right to worry about unitarity. The
solution in gauge theories is as follows: $\eta$-unitarity of~$\Sf$
together with gauge invariance will imply unitarity of the
$\Sf$-matrix on the `physical subspace'.

For the theory defined by~\eqref{eq:summacumlaude}, we have
$$
\eta = \eta_A \ox \eta_{\rm gh} \sepword{on} H = H_A \ox H_{\rm gh}.
$$
We recall $\eta_A$ is given by
$$
\eta_A = \prod_{a=1}^{\dim G}(-)^{\N_{0a}},
$$
where $\N_{0a}$ is the number operator for gauge particles of
$G$-colour~$a$. The gauge potentials $A_\mu^a$ are $\eta$-hermitian.
Grosso modo: we expect the $\eta$-adjoint fields $u^+,\ut^+$ to enter
$T_1$, in order to have $\eta_{\rm gh}$-hermitian quantities. The key
is causality: the latter Krein operator must be defined in a way that
$u^+,\ut^+$ are relatively local to $u,\ut$; we know $u^\7,\ut^\7$ do
not have this property. With all this in mind, we search for the
`good' $\eta_{\rm gh}$. Clearly, it cannot be relatively local itself,
which is tantamount to involve~$\N_{\rm gh}$. A natural guess would be
to take the (already much used) operator:
$$
E := \exp(i\pi\N_{\rm gh}).
$$
However, consider the ghost and antighost number operators:
$$
N_j := \thalf(\N_{\rm gh} + jQ_{\rm gh}),
$$
for $j=1,-1$. They also have integer spectrum. Moreover:
$$
E = (-)^{N_1 + N_{-1}} = (-)^{N_1 - N_{-1}} = (-)^{Q_{\rm gh}},
$$
so $E$ cannot be the right choice. We consider instead
$$
I := (-)^{N_{-1}} = e^{\tihalf\pi(N-Q_{\rm gh})} = T(\sigma_3).
$$
This is indeed a symmetry.  We do have $Ic_j(p)I=jc_j(p)$, and it is
then quickly seen that
$$
Iu^\7I = \ut; \quad I\ut^\7I = u;
$$
so we have locality.  While this is a perfectly sensible solution to
the problem, $T_1$ and~$Q$ are not $I$-hermitian.  One could write
different, equivalent expressions for the terms involving ghosts in
the Lagrangian (see the discussion in the next paragraph); but first
we submit to convention.  Consider then
\begin{align*}
S &= T(U) := T\big(i(\sigma_1 + \sigma_3)/\sqrt2\big) =
T\Big(e^{i\pi(\sigma_1 + \sigma_3)/2\sqrt2}\Big) \sepword{and}
\\
&\eta_{\rm gh} := SIS^{-1} = T(\sigma_1) = i^{N-\Ga}.
\end{align*}
Now we get:
$$
\eta_{\rm gh}c_j(p)\eta_{\rm gh} = c_{-j}(p),
$$
and
$$
u^+ := \eta_{\rm gh}u^\7\eta_{\rm gh} = u; \quad \ut^+ :=
\eta_{\rm gh}\ut^\7\eta_{\rm gh} = -\ut;
$$
together with
$$
T_1^+ = T_1; \qquad Q^+ = Q.
$$

An alternative definition for the ghost contribution in~$T_1$ would be
given by
$$
\thalf f^{abc}\wick:A^a_\mu u^b\overleftrightarrow{\del^\mu}\ut^c:(x)
\sepword{instead of} f^{abc}\wick:A^a_\mu u^b\del^\mu\ut^c:(x).
$$
Both forms differ by a pure divergence term plus a $Q_{\rm
gh}$-coboundary, that is, a term of the form $[Q_{\rm gh}, K]_+$.
Therefore the first one remains gauge invariant.  The choice of it
would allow the use of~$I$ as Krein operator, preserving all the good
properties.  The second one is employed partly for historical reasons.

\smallskip

To conclude, let us comment again on the different behaviour of the
Poincar\'e group representation in the massive and the massless case.
For the former, the representation is always unitary, and commutes
with all charges~$Q(A)$ and transformations~$T(B)$.  Therefore it
is~$\eta$-unitary as well.  However, for the gauge potentials in the
massless case the representation is not unitary, and $\eta_A$ is
introduced for reasons of covariance.

\subsection*{Acknowledgment}

This work was mostly done at the Departamento de F\'{\i}sica Te\'orica
I of the Universidad Complutense, to which I remain gratefully
indebted.  I acknowledge partial support from CICyT, Spain, through
the grant FIS2005--02309.

\end{document}